\documentclass[aps,prd,twocolumn,superscriptaddress,showpacs]{revtex4}

\pdfoutput=1
\usepackage{amsmath}
\usepackage{amsfonts}
\usepackage{amssymb}
\usepackage{graphicx}
\usepackage{epsfig}
\usepackage{subfig}
\usepackage{xcolor}
\usepackage{slashed}

\begin{document}


\title{A Spinodal Solution to Swampland Inflationary Constraints}


\author{R. Holman}
\email{rholman@minerva.kgi.edu}
\author{B. Richard}
\email{brichard@minerva.kgi.edu}
\affiliation{Minerva Schools at KGI,\\1145 Market Street, San Francisco, CA 94103, USA}

\date{\today}

\begin{abstract}
We show how the inclusion of non-perturbative, {\em dynamical} quantum effects on the evolution of the inflaton can allow for an inflationary phase that is both consistent with cosmological constraints {\em and} avoids the problems associated with the so-called swampland conditions. In particular, for concave potentials such as those preferred by the Planck data, spinodal instabilities associated with tachyonic masses for long wavelength modes induce a second round of inflation, which in essence decouples the tree-level potential from the inflationary phase. We illustrate these points using natural inflation as an example.
\end{abstract}

\maketitle

The possibility that not every low-energy effective field theory(EFT) can have a UV completion consistent with string theory has given rise to the idea of the {\em Swampland} \cite{Vafa:2005ui}. This is the collection of EFTs that cannot belong to the string theory landscape and are pathological in some way. Recently, criteria have been posited for an EFT to avoid living in the swampland \cite{Obied:2018sgi} and it has been argued that these criteria could well disrupt the single-field driven inflationary paradigm \cite{Agrawal:2018own,Kinney:2018nny}. Given that inflation provides such an appealing explanation for problems such as the flatness, horizon, and relic problems as well as a source of metric perturbations consistent with observations of cosmic microwave background (CMB), it would be truly unfortunate if single-field inflation were to sink into the muck of the swampland! 

Is this conclusion final? Some ways to avoid it have been discussed in the recent literature. In ref. \cite{Brahma:2018hrd}, the authors argue that single-field inflation and the swampland conjectures could be reconciled if the initial state were different than the standard Bunch-Davies one (some parts of this conclusion are disputed in ref. \cite{Ashoorioon:2018sqb}) while ref. \cite{Das:2018hqy} claims that warm inflation \cite{Berera:1999ws} can be made swampland safe. If the metric perturbations come from a different source than the inflaton field, this can also be used to build a bridge across the swampland for inflation \cite{Achucarro:2018vey,Kehagias:2018uem}.

In this work, we take a different approach to the problem. In all of the discussions of the problems inflation has relative to the swampland criteria, the unspoken assumption is that the dynamics of the inflaton zero mode was completely described by its potential, perhaps with some perturbative quantum corrections added so as to promote it to the quantum effective potential. However, this is certainly not always the case. One instance where this assumption fails drastically is in the case where the tree-level potential has a spinodal regime, i.e. is concave down so that long wavelength modes can be tachyonic. Note that current CMB data exhibits a marked preference for such potentials \cite{Ade:2015lrj,Akrami:2018odb}. In this case, the field dynamics can then be dominated by the behavior of long wavelength fluctuations rather than the dictates of the potential. These effects are non-perturbative and go well beyond those included in the construction of the effective potential \cite{Cormier:1998nt,Cormier:1999ia}. We will argue here that these effects can make inflationary models based on such potentials consistent with the avoidance of the swampland. 

Before we turn to our calculations, let's recapitulate the discussion in ref. \cite{Agrawal:2018own} concerning the swampland conditions. Using the notation in ref. \cite{Kinney:2018nny} we label the conditions as $\mathcal{SC}_1$ and $\mathcal{SC}_2$. The first condition $\mathcal{SC}_1$ constrains field excursions to be less than the Planck scale, i.e. 
\begin{equation}
\label{eq:SC_one}
		\left|\Delta \phi \right| \lesssim C_1 M_{Pl} \sim M_{Pl},
	\end{equation}
with the constant $C_1$ of order unity. This constraint ensures that we can use the EFT, as developed in terms of operators involving powers of $\phi\slash M_{\rm Pl}$ to describe the field behavior consistently. 

The second condition $\mathcal{SC}_2$ restricts the shape of the potential $V(\phi)$ and is, to some extent, the encoding of the fact that it has been impossible to date to find a de Sitter like solution to the string equations of motion that is fully under calculational control. It states that 
\begin{equation}
\label{eq:SC_two}
		M_{Pl}\frac{\left|\nabla_{\phi} V \right|}{V} \gtrsim C_2	
\end{equation}
where $C_2 \backsim \mathcal{O}(1)$.

It is easy to see how these conditions might be roadblocks to constructing a successful single-field inflationary model. The first one precludes large field models \cite{Brown:2015iha} (and while it might seem that models such as those involving axion monodromy \cite{Silverstein:2008sg} could avoid this conclusion, it appears that they too could have problems with the swampland \cite{Blumenhagen:2017cxt}). The second one is even more stringent since it would argue that even small field models could not inflate since $\mathcal{SC}_2$ seems to require that the slow-roll parameter $\epsilon$ be larger than one. In particular the authors of ref. \cite{Kinney:2018nny}, conclude that the bounds on $C_2$ in $\mathcal{SC}_2$ were not consistent with the current allowed values for $n_s$ and $r$. $\mathcal{SC}_2$ ultimately imposes a restriction on allowed values of the tensor-to-scalar ratio, $r$, $r > 8 C_2^2$ \cite{Dias:2018ngv} and given that $C_2$ is $\mathcal{O}(1)$, this is clearly in tension with observational data in which $r$ is predicted to be of less than $0.064$ at the 95\% confidence level \cite{Akrami:2018odb}. 

The effects we are searching for come from non-perturbative growth of fluctuations around the zero mode. In order to control these fluctuations their effects must be resummed and the method we use is the Hartree approximation \cite{Cormier:1998nt,Cormier:1999ia}. To be fair, this is more of a truncation than an approximation, in that there is no systematic way to go beyond it. This is unlike the large $N$ approximation in which we have a control parameter to determine how good the approximation is. However, the Hartree approximation {\em does} recognize the importance of the spinodal line in a dynamical way. Furthermore, whereas rescattering effects could disrupt the approximation, at weak enough coupling, these rates are smaller than the expansion rate during inflation and so can be neglected.

The approximation consists of the replacements:
\begin{eqnarray}
\label{eq:hartree}
\psi^{2n}&\rightarrow& \frac{(2 n)!}{2^n (n-1)!} \langle \psi^2 \rangle^{n-1} \psi^2- \frac{(2 n)!(n-1)}{2^n n!} \langle \psi^2 \rangle^{n} \nonumber\\
\psi^{2n+1}&\rightarrow& \frac{(2 n+1)!}{2^n (n-1)!} \langle \psi^2 \rangle^n \psi,
\end{eqnarray}
where we decompose the full field $\Phi(\vec{x},t)$ as $\Phi(\vec{x},t)=\phi(t)+\psi(\vec{x},t)$. We then further decompose the fluctuations $\psi(\vec{x},t)$ in terms of momentum modes 
\begin{equation}
\label{eq:modedecomp}
\psi(\vec{x},t)=\int \frac{d^3 k}{(2\pi)^3}\ g_{\vec{k}}(t) e^{-i\vec{k}\cdot \vec{x}},
\end{equation}
with $\langle \psi^2 \rangle$ then given by
\begin{equation}
\label{eq:twopoint}
\langle \psi^2 \rangle=\int \frac{d^3 k}{(2\pi)^3}\ \left|g_{\vec{k}}(t)\right|^2.
\end{equation}
While the theory becomes quadratic within the Hartree approximation, the mode equations involve the two-point function, which in turn involves the modes; the interactions are treated via this self-consistency condition.

We consider natural inflation (NI) \cite{Freese:1990rb,Adams:1992bn,Freese:2014nla} as the poster child for models with spinodal instabilities. In natural inflation, the expansion of the universe is driven by an axion $\Phi$, a Pseudo-Nambu-Goldstone Boson (PNGB) of a broken symmetry. The potential for $\Phi$ is given by
	\begin{equation}
		V(\Phi) = \Lambda^4 \left[1+ \cos\left(\frac{\Phi}{f}\right)\right]
		\label{NIPot}
	\end{equation}
where $f$ is the axion decay constant and $\Lambda$ is the mass scale at which strongly coupled interactions generate a potential. The appeal of this model then resides in the presence of a residual shift symmetry which preserves the flatness of the potential from being spoiled by the presence of quantum corrections. As a result, the slow-roll parameters $\epsilon$ and $\eta$ take on values that are consistent with slow-roll inflation. However, in order to satisfy cosmological constraints on the minimum required number of e-folds as well as on the scalar spectral index $n_s$ and the tensor-to-scalar fluctuation ratio $r$ we need $f > M_{Pl}$, where $M_{Pl}$ is the reduced Planck constant. This in turn means that trans-planckian field excursions $\Delta\Phi > M_{Pl}$ (and correspondingly field values greater than $M_{Pl}$) would then occur during inflation which would invalidate the effective field theory approximation that allowed us to only keep the above operators in the theory, rendering the theory non-predictive. Spinodal instabilities have been shown by us to obviate this argument \cite{Albrecht:2014sea}.

To apply the Hartree approximation in eq.\eqref{eq:hartree} to the potential eq.\eqref{NIPot}, we insert the decomposition of the field, Taylor expand the trigonometric functions of the fluctuations, apply the Hartree approximation term by term in this expansion and then resum it. What this yields are the following equations of motion when the theory is embedded in an FRW universe:
\begin{eqnarray}
\label{eq:eqsofmotion}
&&\ddot{\phi}(t) + 3H(t)\dot{\phi}(t) - \frac{\Lambda^4}{f}\exp\left(-\frac{\langle \psi^2 \rangle}{2f^2}\right)\sin\left(\frac{\phi}{f}\right) = 0, \nonumber \\
&&\ddot{g}_k(t) + 3H(t)\dot{g_k}(t) + \nonumber\\
&&\left[\frac{k^2}{a^2(t)}- \frac{\Lambda^4}{f^2}\exp\left(-\frac{\langle \psi^2 \rangle}{2f^2}\right)\cos\left(\frac{\phi}{f}\right) \right] g_k(t) = 0, \nonumber\\
&&H^2(t) = \frac{1}{3 M_{\rm Pl}^2}\left[ \frac{1}{2}\dot{\phi}(t)^2+ \frac{1}{2} \langle \dot{\psi}^2 \rangle+\frac{1}{2} \langle \left(\nabla{\psi}\right)^2 \rangle\right .\nonumber\\
&&\left . +\Lambda^4\left(1+\cos\frac{\phi}{f} \exp\left(-\frac{\langle\psi^2\rangle}{2 f^2}\right)\right) \right].
\end{eqnarray}
These equations show that one of the effects of the growth of the two-point function due to spinodal instabilities is to suppress the effect of the tree-level potential in the equations of motion. It is this suppression that can drive the shift between a potential-driven inflationary phase and a fluctuation-driven one. 

\begin{figure}[ht]
	\centering
	\subfloat[]{
	\includegraphics[scale=0.60]{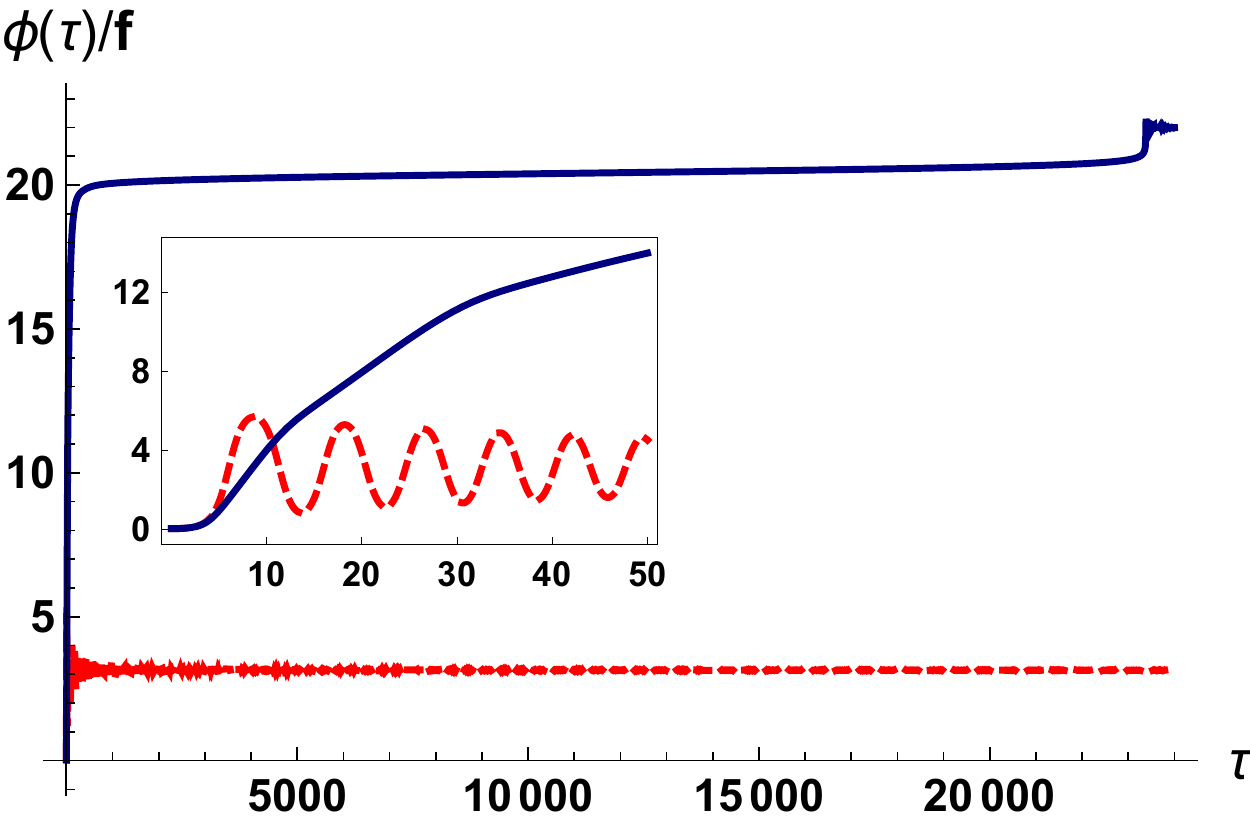} 
	}
	\quad
	\subfloat[]{
	\includegraphics[scale=0.60]{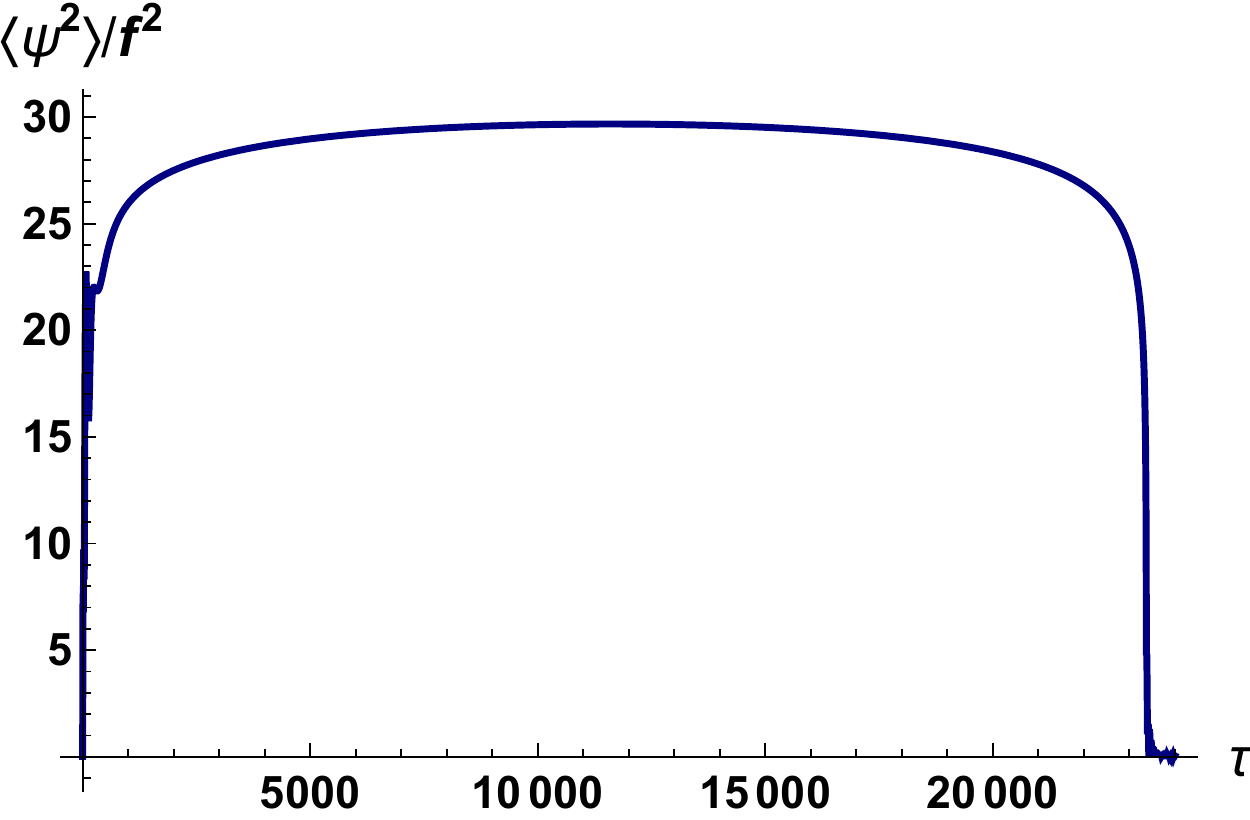}
	}
	\quad
	\subfloat[]{%
	\includegraphics[scale=0.60]{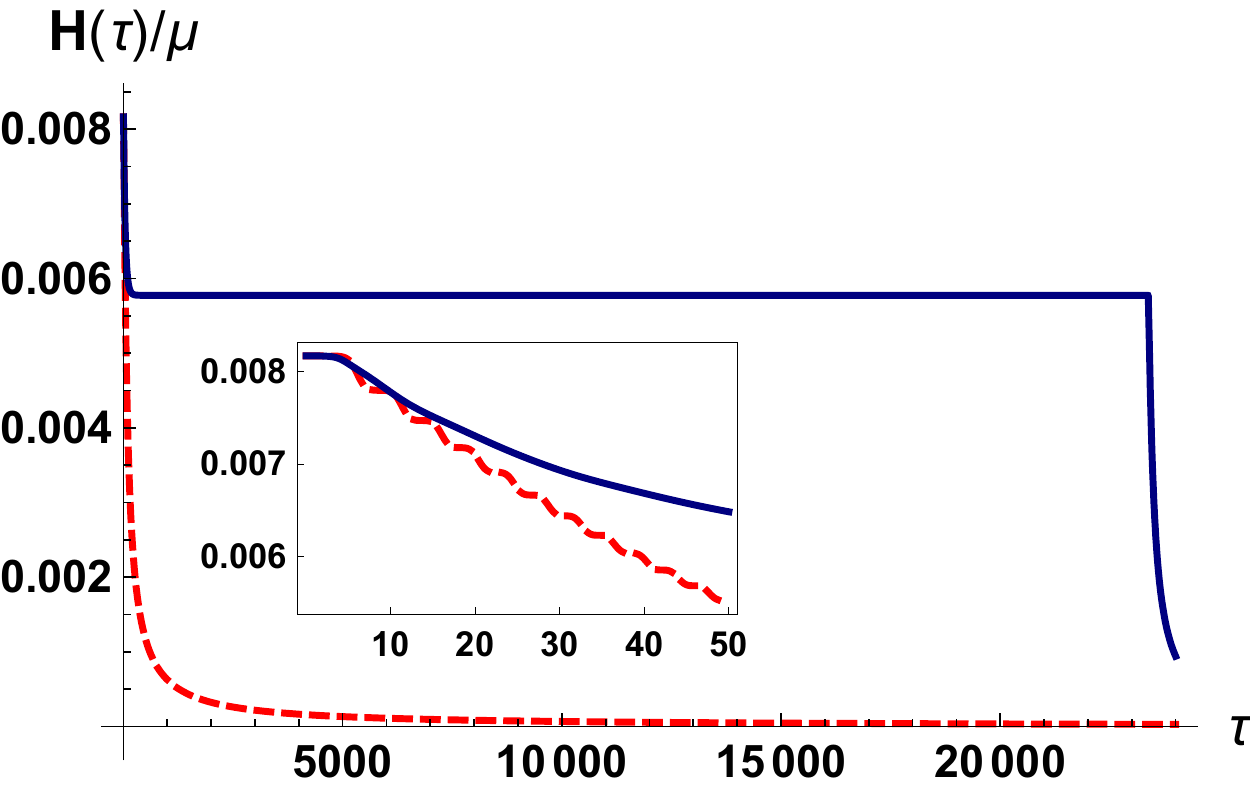}
	}
	\caption{In all the above plots, the quantities are depicted in dashed and red for vanilla inflation and blue for spinodal inflation. (a) The field $\phi(\tau)\slash f$ , (b) The fluctuations $\langle \psi^2 \rangle\slash f^2$, (c) The Hubble parameter $H(\tau)\slash\mu$. The insets show the early time evolution of the first and third quantity.}
	\label{fig:SI}
\end{figure}

Let $\mu = \Lambda^2/f$, $\alpha = {f}\slash{M_{Pl}}$, $\lambda = \Lambda\slash f$, and define $\tau = \mu t$. Fig.\ref{fig:SI} shows the evolution of the system for a choice of parameter values, including the initial value of the inflaton zero mode (we take the initial velocity to vanish) that allows for spinodal instabilities. We see that the dynamics of the field $\phi(\tau)$ is greatly affected by taking the unstable modes into account, with a period during which its value only barely changes as a function of $\tau$. This is due to the dominance of the spinodal modes during that time. On the contrary, in what we call ``vanilla'' inflation where we just use eq.\eqref{NIPot} to drive the zero mode evolution, the inflaton quickly reaches its equilibrium position at $\phi\slash f=\pi$ (the red dashed line in fig.\ref{fig:SI}). The result of $\phi(\tau)$ remaining essentially constant is to induce a second phase of inflation at a lower value for $H$ corresponding to the value of the potential near the spinodal line. This second phase of inflation is no longer potential driven as those features are exponentially suppressed by the growth of $\langle \psi^2 \rangle\slash (2f^2)$, which is clear from eq.\eqref{eq:eqsofmotion}. Instead, the fluctuations run the show. This can clearly be observed in fig.\ref{fig:SI} as the second phase of inflation is in direct correspondence with the domain in time over which $\langle \psi^2 \rangle\slash (2f^2)$ is at its highest.

We see that the cosmological evolution of spinodally enhanced natural inflation is very different than its vanilla counterpart. The question remains: are these changes sufficient to allow for cosmologically consistent inflation that is also swampland safe? We interpret cosmological consistency as being able to generate a sufficient number of e-folds and give rise to a power spectrum consistent with the measured values of the spectral index $n_s$ and the tensor-to-scalar ratio $r$. The swampland safety issue is a more intriguing one in our context. 

As we discussed at the beginning, $\mathcal{SC}_2$ constrains the shape of the potential. However, if the evolution is {\em not} potential driven, then this constraint shouldn't apply. Thus we interpret swampland safety as: during the phase in which the evolution is dominated by the potential, can we choose parameters in the theory and start from initial conditions that are swampland safe {\em and} remain swampland safe until the evolution becomes fluctuation dominated? The answer is yes and we exhibit a set of parameters and initial conditions that satisfy this below. We have not performed an extensive search in the parameter space, so this should be taken in the spirit of an existence proof for such a set of parameters. 

\begin{figure}[ht]
	\centering
	\includegraphics[scale=0.60]{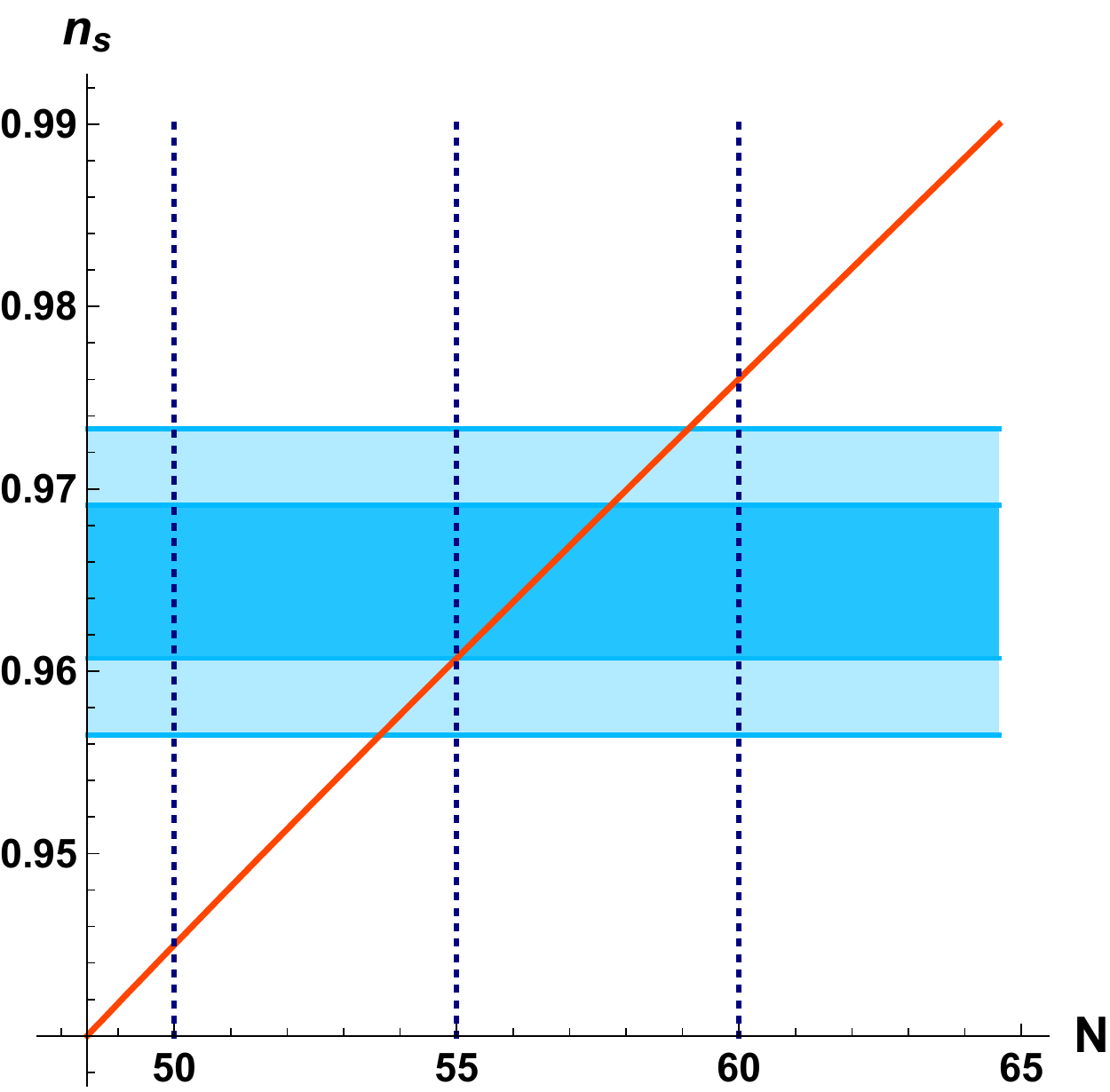} 
	\caption{The spectral index $n_s$ as a function of the number of inflationary e-folds $N$ before the end of inflation. The blue area corresponds to the latest Planck constraints on $n_s$, $n_s = 0.9649 \pm 0.0042$ at $68\%$ CL. The most opaque region corresponds to the $1 \sigma$ region while the least opaque represents how far the $2\sigma$ region extends. It is clear that 55 e-folds before the end of inflation one is well within the $68\%$ confidence interval of the admitted value for $n_s$.}
	\label{fig:ns}
\end{figure}

We take $\alpha=0.01$ corresponding to $f=10^{-2} M_{\rm Pl}$, $\lambda = 1$ so that $\Lambda=f$ and an initial value of $\phi$, $\phi_0=2 \times 10^{-2} f$. With these parameters, the total number of e-folds is $\sim 135$ and we can plot $n_s$ as a function of the number of e-folds $N$ before the end of inflation; this is shown in fig.\ref{fig:ns}. The details of the calculation for $n_s$ and $r$ can be found in ref. \cite{Cormier:1999ia} and ref. \cite{Albrecht:2014sea}. We see from fig.\ref{fig:ns} that our model is well within the $1\sigma$ interval for $N = 55$. Depending how many e-folds of inflation one requires, increasing (decreasing) $\phi_0$ pushes the $N = 60$ ($N = 50$) line within the $68\%$ CL region of $n_s$. The tensor-to-scalar ratio $r$ is exceedingly small, of order $10^{-9}$.

 Are these parameters swampland safe? Clearly $\mathcal{SC}_1$ is satisfied; there is no trans-Planckian excursion by the zero mode or the fluctuations for that matter. To check that $\mathcal{SC}_2$ is not violated we go back to eq.\eqref{NIPot} and see that the second swampland conjecture corresponds to imposing
	\begin{equation}
		\frac{\left|-\sin\left(\frac{\Phi}{f}\right)\right|}{1 + \cos\left(\frac{\Phi}{f}\right)} = \left|\tan\left(\frac{\Phi}{2f}\right)\right| \gtrsim \frac{C_2f}{M_{Pl}} \backsim \frac{f}{M_{Pl}} = \alpha,
		\label{SC2NI}
	\end{equation}

Consequently, $\mathcal{SC}_2$ imposes a constraint on the initial condition for the zero mode $\phi_0$. For our parameter choices, we do satisfy eq.\eqref{SC2NI} while the evolution is potential driven, which by our lights is equivalent to being swampland safe.

To conclude then, we argued here that the swampland safety constraints, as they relate to inflation, are conditioned on there being a potential upon which we have to enforce slow-roll conditions. What we show here is that non-perturbative dynamical effects can make these constraints moot; the system can become fluctuation dominated and the potential is mostly irrelevant during the inflationary evolution. The particular type of non-perturbative effect we've used here, the spinodal instability due to the concavity of the tree-level potential, can be argued to be cosmologically important given that the Planck data prefers such potentials. 

It is worth noting that there is other evidence that non-equilibrium quantum effects can radically modify the potential. Ref. \cite{Boyanovsky:2001va} considered a theory consisting of a scalar field coupled to $N$ fermions and treated zero mode dynamics using the large N approximation. Instead of being driven by the large N effective potential, which is unbounded from below, the dynamics was found to be dominated by the fermion fluctuations which dynamically generated an {\em upright} quartic potential! This example just reiterates our point that statements made about the potential of the theory can be totally divorced from the actual dynamics of the system. At a minimum, one should always take into consideration the full non-equilibrium dynamics of the particle content driving the inflationary period, especially when imposing the swampland conjectures into the mix. 

One should also keep in mind that the Swampland conditions are conjectures. Whether they are valid remains an open question. What we have demonstrated in this work is, assuming their validity, that the conjectures did not preclude a specific category of inflationary models from maintaining its predictive power. 

\bibliography{SpinodalSwamp}

\end{document}